\documentclass[english,aps,prl,twocolumn,superscriptaddress,showpacs]{revtex4}
\usepackage[pdftex]{graphicx}
\DeclareGraphicsExtensions{.pdf,.jpg}
\usepackage{hyperref}
\usepackage{amssymb}
\usepackage[cp1250]{inputenc}
\begin{document}

\title{Quantum effects in linear and non-linear transport of \textsf{T-}shaped
ballistic junction}

\author{J.~Wróbel}
%\email{wrobel@ifpan.edu.pl}
\author{P.~Zagrajek}
\author{M.~Czapkiewicz}
\affiliation{Institute of Physics, Polish Academy of Sciences, al Lotników 32/46,
02-668 Warszawa, Poland}

\author{M.~Bek}
\affiliation{Institute of Molecular Physics, Polish Academy of Sciences, ul. M. Smoluchowskiego 17, 60-179 Poznań, Poland}

\author{D. Sztenkiel}
\author{K.~Fronc}
\affiliation{Institute of Physics, Polish Academy of Sciences, al Lotników 32/46,
02-668 Warszawa, Poland}

\author{R.~Hey}
\author{K.~H.~Ploog}
\affiliation{Paul Drude Institute of Solid State Electronics, Hausvogteiplatz
5-7, D-10117 Berlin, Germany}

\author{B.~R.~Bułka}
\affiliation{Institute of Molecular Physics, Polish Academy of Sciences, ul. M.
Smoluchowskiego 17, 60-179 Poznań, Poland}

\date{\today}
\begin{abstract}
We report low-temperature transport measurements of
three-terminal \textsf{T}-shaped device patterned from
GaAs/Al$_{x}$Ga$_{1-x}$As heterostructure. We demonstrate
the mode branching and bend resistance effects predicted by
numerical modeling for linear conductance data. We show also that the
backscattering at the junction area depends on the wave
function parity. We find evidence that in a non-linear transport
regime the voltage of floating electrode always \emph{increases} as
a function of\emph{ push-pull} polarization. Such anomalous effect occurs for the
symmetric device, provided the applied voltage is less than the Fermi energy in equilibrium.
\end{abstract}

% insert suggested PACS numbers in braces on next line
\pacs{73.21.Nm, 73.23.Ad, 85.35.Ds}
% insert suggested keywords - APS authors don't need to do this
\keywords{quantum transport, three-terminal branch
switches, linear and nonlinear effects.}
\maketitle
%%%%%%%%%%%%%%%%%%%%%%%%%%%%%%%%%%%%%%%%%%%%%%
Recently, nanotechnology advances have led to a growing interest in
electrical transport properties of the so-called three-terminal ballistic
junctions (TBJs). As the name indicates, such structures consist of
three quantum wires connected via a ballistic cavity to form a \textsf{Y}-shaped
or \textsf{T}-shaped current splitter. One motivation is that in principle
such systems can operate at high speed with a very low power consumption.
Therefore, interesting and unexpected nonlinear transport characteristics
of TBJs are intensively investigated due to possible applications
as high frequency devices or logic circuits\cite{Xu2005,Worschech2005}. 

Another reason for the increased number of studies devoted to TBJs
are quantum mechanical aspects of carrier scattering, which dominate
at low temperatures in the linear transport regime. This applies especially
to \textsf{T}-shaped splitters. For example, it is expected that a
\textsf{T}-branch switch, made of materials with a significant spin-orbit
interactions, can act as an effective spin polarizer \cite{Kiselev2001}.
Also, for such geometry an ideal splitting of electrons from a Cooper
pair is expected, provided the lower part of the letter \textsf{T
}is made of a superconducting material \cite{Bednorz2009}. Both effects
rely very strongly on the perfect shape of the devices and high enough
transparency of individual wires. Unfortunately, experimental data
available for the lithographically perfect \textsf{T}-branch junctions
are limited mostly to a non-linear transport regime \cite{Wallin2006-Spanheimer2009}.
Quantum linear transport is usually studied for less symmetric structures,
typically consisting of short point contact attached to a side wall
of a wider channel \cite{Usuki1995-Ramamoorthy2007}.

\begin{figure}
\includegraphics{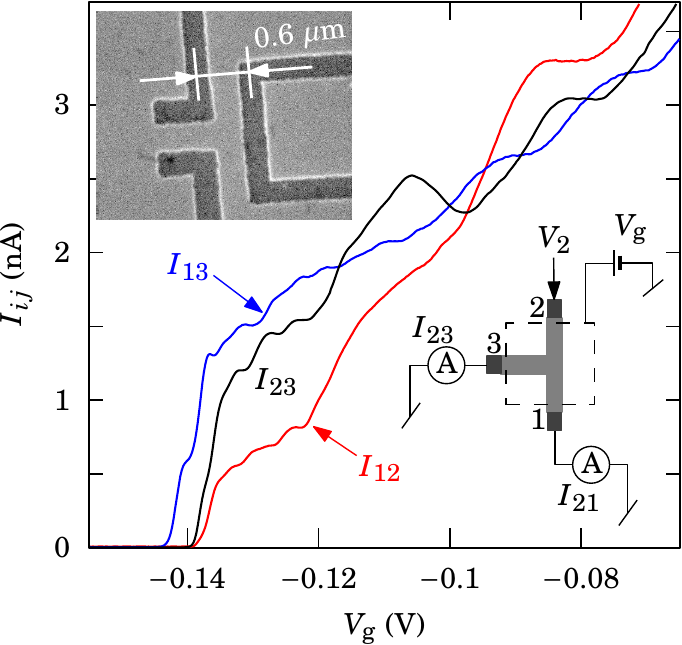} 
\caption{(Color online) Currents $I_{ij}$ vs gate voltage $V_{\mathrm{g}}$ at temperature
$T\approx0.3$ K. $I_{ij}$ is defined as current flowing from contact
$j$ when voltage $V_{i}$ is applied to terminal $i$ (see the measurement
scheme). Upper inset shows scanning electron
micrograph of the \textsf{T}-junction device, top metal gate is not
visible here. \label{fig:Ijk}}

\end{figure}

In this work we report on fabrication and low temperature transport
measurements of \textsf{T}-shaped three-terminal devices, for which
we take a special care to preserve the perfect symmetry and reduce
the geometrical disorder. By comparing our data to conductance modeling
by the recursive Green-function method, we find out that quantum effects
dominate up to source-drain voltages equal to the Fermi energy. In
particular, we show that the non-linear response of symmetric TBJ
behaves in a non-classical way and is highly tunable with carrier
density.

The three-terminal ballistic junctions are made of a GaAs/AlGaAs:Si
heterostructure with electron concentration $n_{2D}=2.3\times10^{11}$~cm$^{-2}$
and carrier mobility $\mu=1.8\times10^{6}$~cm$^{2}$/Vs. The interconnected
wires of equal length $L=0.6$~$\mu$m and lithographic width $W_{\mathrm{lith}}=0.4$~$\mu$m
are patterned by \emph{e}-beam lithography and shallow-etching techniques
to form a \textsf{T}-shaped nanojunction (see inset to Fig.\ref{fig:Ijk}).
The physical width of all branches is simultaneously controlled by
means of a top metal gate which is evaporated over the entire structure.
The differential conductances have been measured in a He-3/He-4 dilution
refrigerator, by employing a standard low-frequency lock-in technique.
We have also studied non-linear transport in the typical for TBJs,
so-called \emph{push-pull} bias regime, when equal but opposite in
sign $dc$ voltages are simultaneously applied to the opposite input
contacts.

The application of a metal gate over the active region of the device
helps to symmetrize transmission coefficients by smoothing the confinement
potential \cite{Liu1994}. Nevertheless, even a perfectly shaped and
gated junction may remain disordered at low electron densities, when
screening effects are weak. Figure \ref{fig:Ijk} shows linear currents
flowing from each of three terminals for negative gate voltages close
to the threshold regime. The data indicate clearly that there is a weak
asymmetry between contacts -- channels open at slightly different
$V_{\mathrm{g}}.$ Additionally, small reproducible wiggles are visible
above threshold voltage. All investigated structures show similar
behavior and we attribute it to the presence of \textit{quasi-}localized
states, formed in the central part of the device. In this paper we
present data for the sample which has a lowest disorder and highest
degree of symmetry.

\begin{figure}
\includegraphics{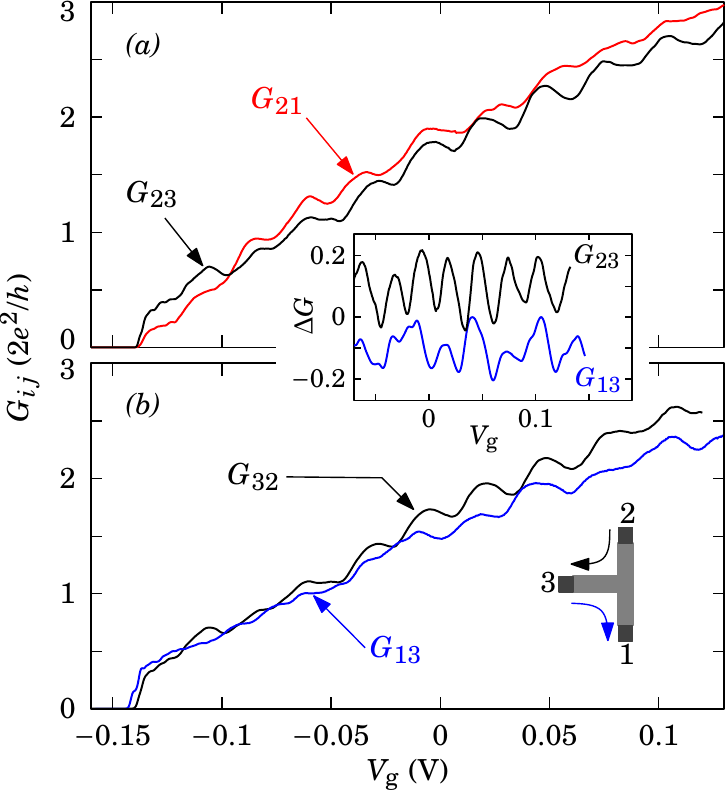} 
\caption{(Color online) $G_{ij}=I_{ij}/V_{i}$ plotted vs gate voltage at $T\approx0.3$~K.
(a) $G_{23}$ and $G_{21}$. (b) $G_{32}$ and $G_{13}$, here
both conductances involve transmission to side terminal $3$. Inset:
comparison between $G_{23}$ and $G_{13}$ oscillations, a smooth
backgrounds have been removed from the original data ($\Delta G$
is in $2e^{2}/h$ units, $V_{\mathrm{g}}$ is in volts). \label{fig:Gjk}}

\end{figure}

Although channel $2\rightarrow1$ opens last, at higher electron densities
$I_{12}$ is larger than $I_{23}$ and $I_{13}$, as predicted by
Baranger \cite{Baranger1990} for the ideal \textsf{T}-shaped quantum
splitter. Figure \ref{fig:Gjk} presents the conductances $G_{ij}$
as a function of gate voltage up to $+0.12$ V. For $V_{\mathrm{g}}>-0.05$
V the regular oscillations corresponding to the successive population
of electric sub-bands in each of the three terminals are visible.
Since magnetic field is zero, we expect $G_{ij}=G_{ji}$ and this
is indeed observed in the experiment. For example, curves $G_{23}$ and
$G_{32}$ are almost identical. Larger differences
are noticed for $G_{13}$ and $G_{23}$ curves which should be equal
for the perfectly shaped device. Relevant data are presented in the
inset to Fig.\ref{fig:Gjk} where oscillating parts of $G_{23}$ and
$G_{13}$ are compared. On average $G_{13}$ is smaller and oscillate
less regularly than $G_{23}$. Nevertheless, maxima and minima on
both curves are close to each other and for $V_{\mathrm{g}}>0.05$
V they oscillate exactly in phase. It means that starting from a disordered
structure at the threshold voltage, for $V_{\mathrm{g}}\gtrsim0$
the device becomes more symmetrical and experimental data can be compared
with the theory of ballistic transport.

\begin{figure}
\includegraphics{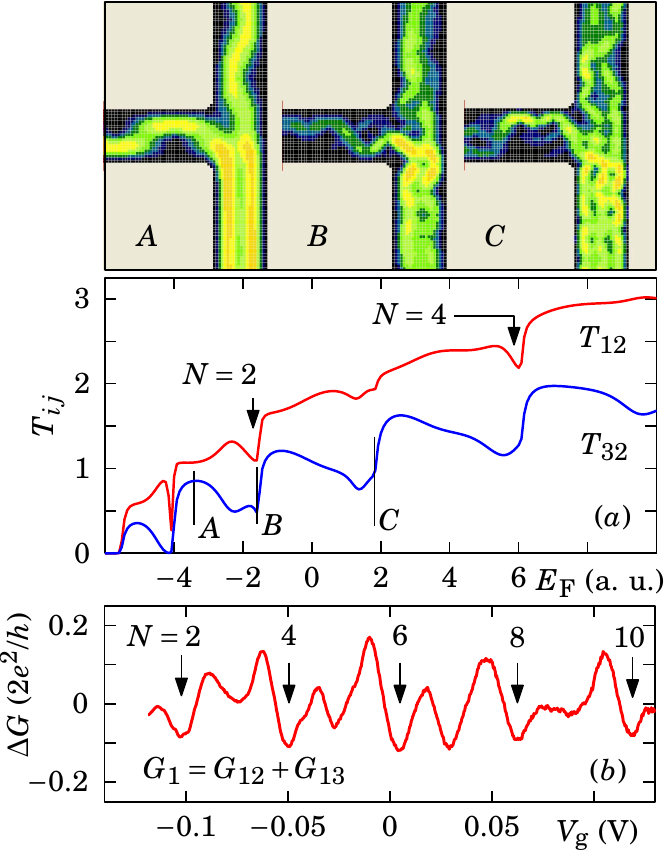} 
\caption{(Color online) (a) Local current intensity (upper panel) and transmission
coefficients $T_{ij}$ vs Fermi energy $E_{F}$ (below). Lines $A$,
$B$ and $C$ mark energy values for which the local current densities
have been calculated. Black color in density plot corresponds to zero
current and bright areas to maximal current intensity. (b)\emph{ }Conductance
$G_{1}=G_{12}+G_{13}$ vs gate voltage, $T\approx0.3$~K. Only oscillating
part is shown, a smooth background has been removed. Arrows on both
subfigures indicate backscattering at even mode numbers. \label{fig:teo-Tij}}

\end{figure}

We model TBJ by three semi-infinite strips of ``atoms'' and the
square coupling region. Calculations have been performed at temperature
$T=0$, using a tight-binding approach and a recursive Green functions
technique \cite{Bulka2009}. To determine a local current intensity
inside the junction we have incorporated parts of each wire to the
coupling region and used a newly developed, so-called knitting algorithm
\cite{Kazymyrenko2008}. Results of this modeling are presented in
Fig. \ref{fig:teo-Tij}(a). Transmission coefficients $T_{ij}$ between
$j$-th and $i$-th electrode are calculated for disorder free and
symmetric device with\emph{ rounded corners} in the coupling region.
Note that the value of $T_{21}$ increases almost monotonically as a function
of energy, whereas $T_{32}$ oscillates strongly. This is the co-called
\textit{bend resistance effect}. $T_{32}$ reaches maximum when the
upper, just populated sub-band, is almost fully transmitted to the
terminal $3$ (see intensity plot $A$). For higher kinetic energies,
however, coupling becomes weaker and as a result $T_{32}$ decreases,
leading to the non-monotonic behavior as a function of Fermi energy
$E_{\mathrm{\mathrm{F}}}$. 

Presented calculations are consistent with the experimental data obtained
at electron densities high enough. For $V_{\mathrm{g}}>0$ the curve $G_{21}$
is similar to $T_{21}$ and rather smooth as compared to $G_{23}$,
which (like $T_{23}$) shows deeper minima due to the bend resistance
effect (see Fig. \ref{fig:Gjk}). Note also, that calculated energy
dependence of transmission coefficients differ for odd and even channel
numbers. For example, the backscattering for $N=2$ and $N=4$ channels
is stronger, as indicated with arrows in Fig. \ref{fig:teo-Tij}.
This effect was already predicted for a perfect \textsf{T} coupler
\cite{Baranger1990} and is apparently enhanced by the rounding of
the ``corners'' in a junction area. For even parity modes electron
has high probability density at the center of the device and therefore
is more likely transmitted (to see this compare density plots $B$
and $C$). We believe that such conductance dependence on wave function
parity is also observed in the experiment. It is especially well resolved
for the total conductance $G_{1}=I_{1}/V_{1}=G_{12}+G_{13}$. Relevant
data are presented in Fig.~\ref{fig:teo-Tij}(b).

Next we consider the measurement scheme where stub terminal ($3$)
acts as a floating voltage probe ($I_{3}=0$). For a classical device
we have $V_{3}=(V_{1}-V_{2})/2$. This simple formula should be modified
for ballistic transport, where it takes form $V_{3}/V_{1}=T_{31}/(T_{31}+T_{32})$
with $V_{2}=0$ for simplicity. If $T_{31}=T_{32}$ then classical
result $V_{3}/V_{1}=1/2$ is recovered.

\begin{figure}
\includegraphics{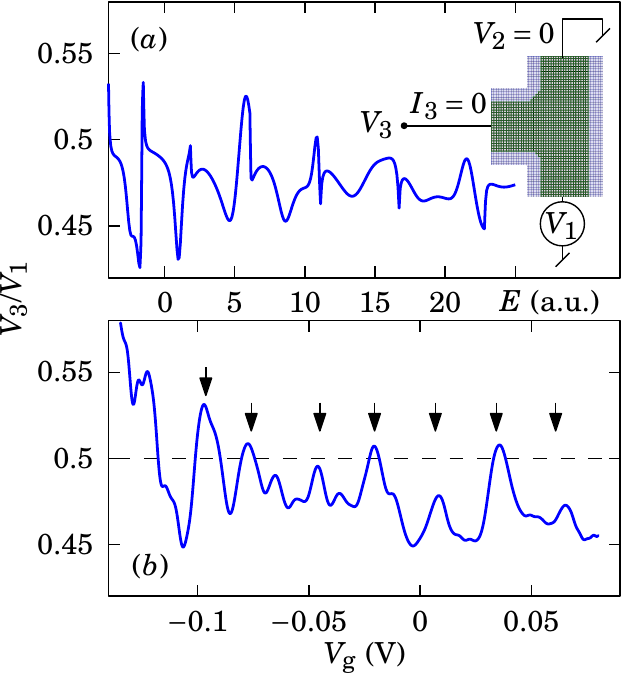}

\caption{(Color online) (a) $V_{3}/V_{1}$ ratio vs energy, calculated for a device with 
asymmetrically rounded corners in the coupling region (see inset).
(b) $V_{3}/V_{1}$ data obtained as a function of gate voltage at $T\thickapprox0.2$
K for $V_{1}=50$ $\mu$V (measurement scheme is shown above). Arrows
correspond to minima on $G_{23}$ curve. \label{fig:V3-lin}}

\end{figure}

Conductance data shown in Fig. \ref{fig:Gjk}(b) indicate that on
average $G_{31}$ is smaller than $G_{32}$. Therefore, to imitate
the real sample, we rounded the junction ``corners'' of a model device
in such a way that $T_{31}<T_{32}$. The shape of the coupling area
and results of calculations are shown in Fig. \ref{fig:V3-lin}(a).
Ratio $V_{3}/V_{1}$ is on average below $1/2$ but oscillates as
energy increases. Very similar dependence is observed in the experiment.
The measured value of $V_{3}/V_{1}$ ratio reaches maximum, each time
a new one-dimensional level becomes occupied. Interestingly, theory
also predicts the occurrence of additional asymmetric and very narrow
resonances when a new conduction channel opens to transport in stub
terminal. They are probably related to the so-called Wigner singularities,
which exist when the energies of quantized levels in a side probe
differ from those in the rest of the device\cite{Bulka2009}. Similar
features are also visible in the experiment, especially for $-0.1<V_{\mathrm{g}}<0$,
but their possible connection to Wigner resonances requires further
studies.

\begin{figure*}
\includegraphics{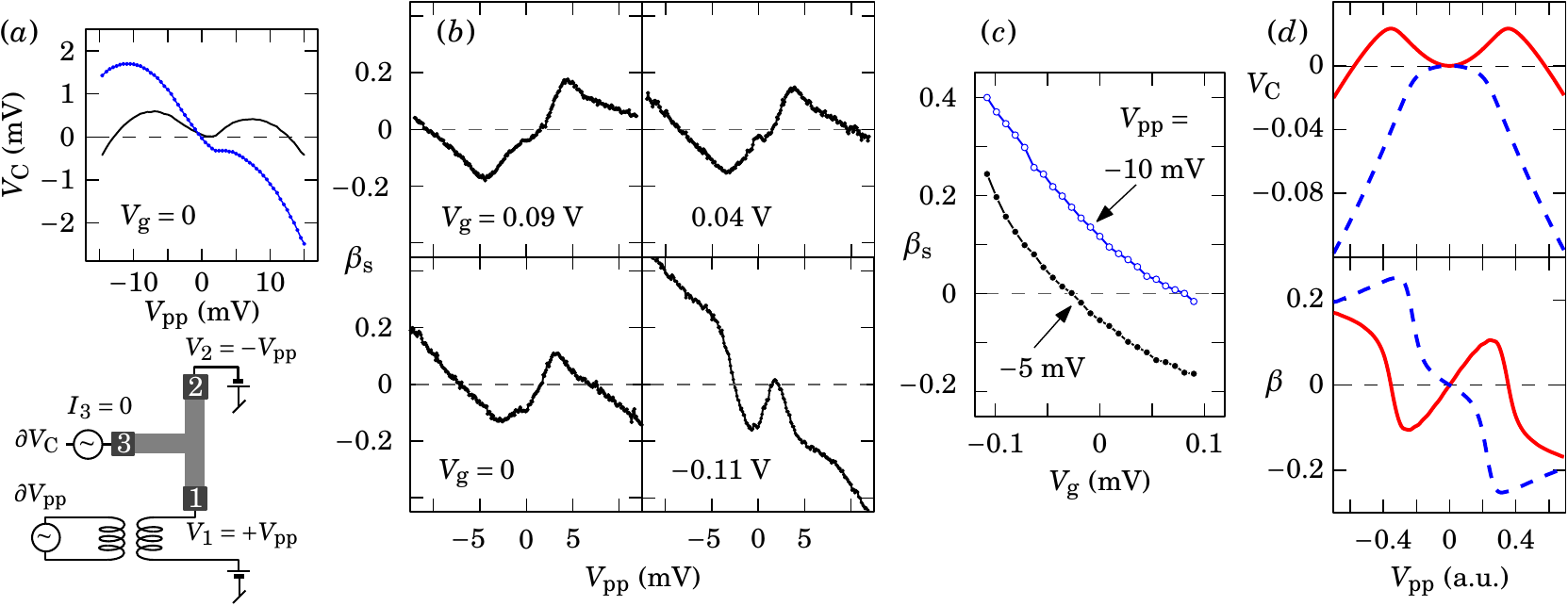}

\caption{(Color online) (a) Stub voltage $V_{\mathrm{C}}$ vs push-pull polarization $V_{\mathrm{pp}}$
at $V_{\mathrm{g}}=0$ (dotted line). The same data with a linear
trend removed are also shown (solid line). Below: experimental setup;
small \emph{ac} voltage ($50$ $\mu$V) is inductively coupled to
$V_{\mathrm{pp}}$ , $\beta=\partial V_{\mathrm{C}}/\partial V_{\mathrm{pp}}$
is measured directly using a low-frequency lock-in technique. (b)
Variation in $\beta_{\mathrm{s}}=\beta-\beta_{\mathrm{a}}$ with the
applied $V_{\mathrm{pp}}$ for $V_{\mathrm{g}}$ of $0.09$ , $0.04$,
$0$ and $-0.11$ V; here $\beta_{\mathrm{a}}$ is the mean value
of $\beta$ and equals $-0.18$, $-0.15$, $-0.12$, and $0.01$
respectively. (c) Variation in $\beta_{\mathrm{s}}$ with gate voltage
for $V_{\mathrm{pp}}$ of $-10$ and $-5$ mV. All experimental results
at $T=0.8$ K. (d) Nonlinear transport data calculated for an ideal
\textsf{T}-shaped junction. Solid line: $E_{\mathrm{F}}=-1.55$, $\partial T_{\mathrm{31}}/\partial E_{\mathrm{F}}<0$.
Dashed line: $E_{\mathrm{F}}=-1.15$, $\partial T_{\mathrm{31}}/\partial E_{\mathrm{F}}>0$.
$E_{\mathrm{F}}$, $V_{\mathrm{C}}$ and $V_{\mathrm{pp}}$ in arbitrary units. 
\label{fig:V3-nolin}}

\end{figure*}

Now let us turn to the non-linear transport regime where the probabilities
of transmission from input terminals to a floating contact may differ,
even for a perfect device. In such case, when $V_{1}$ is large enough
and positive, then $V_{3}/V_{1}$ is \emph{less} then $1/2$. Equivalently,
if $V_{1}=V_{\mathrm{pp}}$ and $V_{2}=-V_{\mathrm{pp}}$ (\emph{push-pull}
bias regime) then $V_{3}=V_{\mathrm{C}}$ is \emph{always negative},
as it was predicted in \cite{Xu2001} and then proved experimentally
\cite{Shorubalko2001-Worschech2001}. Using the quantum scattering
approach Csontos and Xu \cite{Csontos2003} extended the calculation
range to a low temperature regime. They showed that $V_{\mathrm{C}}$
may be also\emph{ positive}, provided $\partial T_{\mathrm{31}}/\partial E_{\mathrm{F}}=\partial T_{\mathrm{32}}/\partial E_{\mathrm{F}}<0$
and $kT\ll E_{\mathrm{F}}$. To our knowledge, however, the predictions
of Ref. \cite{Csontos2003} have not been confirmed experimentally.

Figure \ref{fig:V3-nolin}(a) shows measurement schematics and corresponding
$V_{\mathrm{C}}$ data obtained when $|V_{\mathrm{pp}}|<15$ mV. $V_{\mathrm{C}}$
is not a symmetric function of $V_{\mathrm{pp}}$, yet above a certain
threshold, data --- as expected --- bend towards negative values of
$V_{\mathrm{C}}$. Such behavior is often observed in experiments
\cite{Shorubalko2001-Worschech2001} because $T_{31}\neq T_{32}$ due to imperfections
which are always present in the real devices. Apart from such asymmetry,
however, data reported here behave in an \emph{anomalous} way. When
a linear trend has been removed, $V_{\mathrm{C}}$ first \emph{increases}
with $|V_{\mathrm{pp}}|$, and then goes down reaching maximum at
$\sim7$ mV. To investigate this effect in more detail we have used
a modulation method to measure the \emph{switching parameter} $\beta=\partial V_{\mathrm{C}}/\partial V_{\mathrm{pp}}$
directly with a better voltage resolution. 
Figure \ref{fig:V3-nolin}(a)
explains the measurement idea and Fig. \ref{fig:V3-nolin}(b) shows
values of parameter $\beta_{\mathrm{s}}=\beta-\beta_{\mathrm{a}}$ as
a function of $V_{\mathrm{pp}}$ for a different gate voltages. Here 
$\beta_{\mathrm{a}}$ is the mean value of switching parameter calculated at
each $V_{\mathrm{g}}$ for $|V_{\mathrm{pp}}|<15$ mV. Subtracting $\beta_{\mathrm{a}}$ is
equivalent to removing a linear trend from the \emph{dc} data and therefore
reduces the influence of the $T_{31}$ vs $T_{32}$ asymmetry.

To compare the experimental findings with theory we calculated $V_{\mathrm{C}}$
and $\beta$ for an ideal \textsf{T}-shaped junction from the energy
dependence of a transmission coefficients. Results are consistent
with the explanation of Xu \cite{Xu2001}, as it follows from Fig.
\ref{fig:V3-nolin}(d). If $\partial T_{\mathrm{31}}/\partial E_{\mathrm{F}}<0$
then $V_{\mathrm{C}}$ \emph{increases} with $|V_{\mathrm{pp}}|$
and $\beta$ has a positive slope in this voltage range. When $\partial T_{\mathrm{31}}/\partial E_{\mathrm{F}}>0$
stub voltage is negative and switching parameter behaves ``normally''.
Interestingly, when experimental $V_{\mathrm{C}}$ data are compared
to linear conductance $G_{3}=G_{31}+G_{32},$\emph{ no }such correlation
can be found. For example at $V_{\mathrm{g}}=0$, $0.04$, and $0.09$
V, derivative $\partial G_{\mathrm{3}}/\partial V_{\mathrm{g}}$ is
negative, positive and approximately zero, but switching parameter
does not change its shape and sign as would be expected from modeling.
Results indicate that an anomalous data range, where $\beta$ has
a positive slope, always exists --- only its width decreases with
$E_{\mathrm{F}}$. This fact can be used to tune switching parameter
with the gate voltage. Figure \ref{fig:V3-nolin}(c) shows $\beta_{\mathrm{s}}$
as a function of $V_{\mathrm{g}}$ for the two values of $V_{\mathrm{pp}}$.
Remarkably, not only amplitude but also the \emph{sign} of $\beta_{\mathrm{s}}$
can be changed. We conclude that the behavior of $V_{\mathrm{C}}$
in Fig. \ref{fig:V3-nolin} cannot be explained by a single particle
transmission approach. Probably, as suggested in \cite{Buettiker2003},
the non-linear transport regime requires a self-consistent calculations. 

In summary, we have shown that linear transport in \textsf{T}-shaped
ballistic junction can be successfully described by quantum scattering
effects and weak disorder in a cavity area.  We have
shown for the first time, that stub voltage can \emph{increase} as
a function of \emph{push-pull} polarization in a non-linear transport
regime, however, the energy dependence of such non-equilibrium
effect is inconsistent with the standard single-particle picture of
electron transmission. Nevertheless, novel applications of symmetric TBJ structure, 
for example as the component of a multilogic device, are still possible.
\begin{acknowledgments}
This work was funded by grant No. 107/ESF/2006 and 
MNiSW projects N202/103936 and N202/229437.
\end{acknowledgments}
%%%%%%%%%%%%%%%%%%%%%%%%%%%%%%%%%
\bibliographystyle{apsrev}
%\bibliography{D:/Literature/BIB/TY}

\end{document}